\documentclass[12pt]{article}
\usepackage{graphicx}
\usepackage{amsmath}

\newcommand\pubnumber{SNSN-XXX-YY}
\newcommand\pubdate{\today}

\def\uofm{Physics Department\\
University of Michigan, 500 S State St, Ann Arbor, MI 48109}

\def\Title#1{\begin{center} {\Large #1 } \end{center}}
\def\Author#1{\begin{center}{ \sc #1} \end{center}}
\def\Address#1{\begin{center}{ \it #1} \end{center}}

\newcommand\pubblock{\rightline{\begin{tabular}{l} \pubnumber\\
         \pubdate  \end{tabular}}}
\newenvironment{Abstract}{\begin{quotation}  }{\end{quotation}}
\newenvironment{Presented}{\begin{quotation} \begin{center} 
             PRESENTED AT\end{center}\bigskip 
      \begin{center}\begin{large}}{\end{large}\end{center} \end{quotation}}
\def\Acknowledgements{\bigskip  \bigskip \begin{center} \begin{large}
             \bf ACKNOWLEDGEMENTS \end{large}\end{center}}
\newcommand{\kl}{\mbox{$\rm{K}_L$}}
\newcommand{\klpinn}{\mbox{$\rm{K}_L \rightarrow \pi^0 \nu \bar{\nu}$}}
\newcommand{\kcharged}{\mbox{$\rm{K}^{\pm} \rightarrow \pi^{\pm} \nu \bar{\nu}$}}
\newcommand{\kpimumu}{\mbox{$\rm{K}^{\pm} \rightarrow \pi^{\mp} \mu^{\pm} \mu^{\pm}$}}
\newcommand{\mue}{\mbox{$\mu \rightarrow \rm{e} $}}
\newcommand{\meg}{\mbox{$\mu^+ \rightarrow \rm{e}^+\gamma$}}
\newcommand{\muN}{\mbox{$\mu + \rm{N} \rightarrow \rm{e} + \rm{N^\prime}$}}

\begin{document}
\begin{titlepage}
\pubblock

\vfill
\Title{Rare Decays in Kaons and Muons}
\vfill
\Author{ Monica Tecchio}
\Address{\uofm}
\vfill
\begin{Abstract}
This paper summarizes the status of selected rare decay experiments in kaon and muon physics.
\end{Abstract}
\vfill
\begin{Presented}
XXXIV Physics in Collision Symposium \\
Bloomington, Indiana,  September 16--20, 2014
\end{Presented}
\vfill
\end{titlepage}

\section{Introduction}

The standard model (SM) of electroweak interactions has been incredibly successful but it is likely to be the low-energy limit of a more fundamental theory which we will refer to as new physics (NP).
There are two main approaches to search for new physics: first is to study physics processes that cannot proceed at tree level but are dominated by loops, as NP particles can show up virtually in these loops; the second  is to study physics processes that imply violation of SM conservation laws, like lepton flavor violation (LFV) or lepton number violation (LNV).
These second class of processes are by nature very rare and thus very sensitive to tree level or higher order NP contributions.

Among rare processes, decays involving kaons and muons are particularly attractive because they can be generated in high intensity beams which give access to large statistic samples; and because they have simple decay topologies, and thus clean experimental signatures.
Two experimental approaches are used:  either find observables for which the SM predictions are very accurate and which can be measured with high precision; or just look for highly improbable final states. In the first case, NP signal can be extracted if any deviation from the prediction is observed; in the second, NP arises naturally as the source of any positive observation. In both cases, there is the need for state-of-the-art detectors.

This paper will report on few selected rare kaon and muon searches which have either been pursued recently or are in the planning stage. Section~\ref{sec:kpinn} reports on the kaon golden decays, namely on a new limit for the branching ratio of $\klpinn$ decays by the KOTO experiment at JParc and on the status of the observation of  the $\kcharged$ decay pursued by NA62 at CERN. Section~\ref{sec:kpimumu} reports a recent limit on the $\kpimumu$ decay from the analysis of data taken by the NA48/2 experiment at CERN. Section~\ref{sec:raremuon} presents the status of the present and future experimental sensitivity for $\mu \rightarrow e$ searches: in particular the new limit on $\meg$ decay by the MEG experiment at PSI and the prospectives for $\mue$ conversion detection by the Mu2e experiment at Fermilab. 

\section{The Kaon Golden Decays}\label{sec:kpinn}

\begin{figure}[htb]
\centering
\includegraphics[height=1.5in]{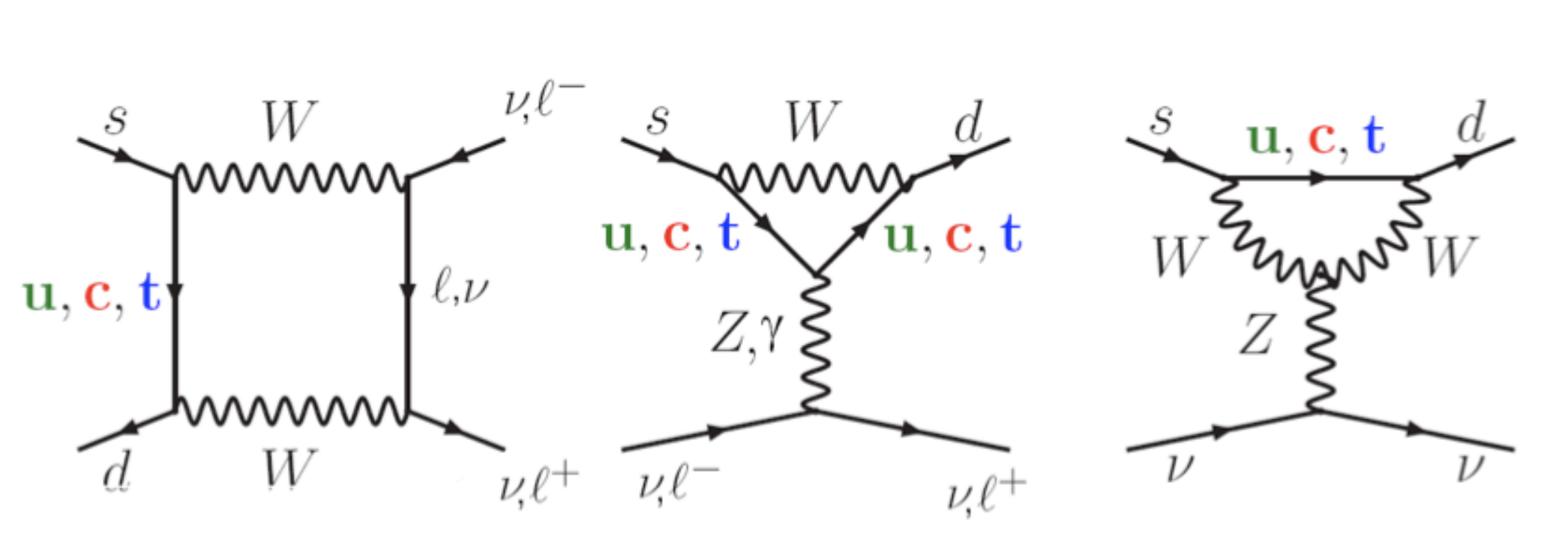}
\caption{From left to right: box diagram and penguin diagram contributions to the kaon golden modes.}
\label{fig:kpinn}
\end{figure}

The two rare kaon decays $\klpinn$ and $\kcharged$ are flavor changing neutral current processes, forbidden at tree level and dominated by one loop diagrams (see Figure~\ref{fig:kpinn}).
The absence of charged leptons in the final states makes the long distance contributions to these decays very small. In addition to that, the relevant hadronic operator can be extracted from the well measured  $\rm{K}^+ \rightarrow \pi^0 \rm{e}^+ \nu$ decay. 
These properties lead to a very clean theoretical prediction for their rates. The next-to-leading order calculations, including full two-loop electroweak corrections to the top quark, predicts B($\klpinn$) = 2.43(39)(6) $\times 10^{-11}$ and B($\kcharged$) = 7.81(75)(29) $\times 10^{-11}$~\cite{Brod_2011} where the first error is parametric and the second is the remaining theoretical uncertainty.
These decays also provide input to the precise determination of the CKM unitarity triangle as well as tests of the Minimal Flavor Violation (MFV) extension of the SM. The wealth of fundamental physics parameters these processes give access to gained them the name of "golden" kaon decays. 

The most precise direct measurement of the charged channel comes from the E787/E949 experiments at Brookhaven: B($\kcharged$) =$(1.73 + 1.15 - 1.05) \times 10^{-10}$~\cite{E787_2009}.
For the neutral decay, the best limit comes from the E391a experiment at KEK:  
B($\klpinn < 2.6 \times10^{-8})\,@\,90\%$ confidence level (CL)~\cite{E391a_2010}.
Several NP scenarios predicts sizable deviations from the SM with a various degree of correlation between the two kaon decays~\cite{Straub_2011} so precise independent measurements of the two rates can help distinguish between different NP models.

\subsection{The KOTO Experiment}

The KOTO collaboration has reported a limit on the $\klpinn$ branching ratio after its first physics run in May 2013 at the 30~GeV/c proton beam in JPARC, Japan. The experiment~\cite{koto} builds on the experimental strategy pioneered by the E391a experiment: a clean narrow $\kl$ beam extracted off-axis to lower the hadron halo momentum to below the $\eta$ production threshold; precisely shaped collimators to minimize halo particles; a highly segmented calorimeter made of about 2800 Cesium Iodide (CsI) crystal; and an hermetic veto system sensitive to both extra photons and/or charge particles in $\kl$ decays. KOTO Phase-I aims at a single event sensitivity (SES) of $9 \times 10^{-12}$ which should result in the first observation of $\klpinn$ at SM predicted rates. A Phase-II upgrade 
is also planned with the goal of observing 100 events for a 10\% measurement.

The experimental technique is to fully reconstruct the $\pi^0$ from the two photons detected in the calorimeter:  using the $\pi^0$ mass as a constraint one can get the photon opening angle. With the further assumption that the $\kl$, and hence the $\pi^0$, vertex is a point along the beam line, one can calculate the pion transverse momentum. The $\klpinn$ final state is isolated from the background by selecting events where the $\pi^0$ has a non-negligible transverse momentum because of the presence of the two neutrinos in the final state.

Despite KOTO first physics run lasted only 100 h (due to a radiation accident) with a beam power of
24 kW, or 10\% of the design intensity, it reached a preliminary SES of (1.29$\pm 0.01_{\rm{stat}} \pm 0.17_{\rm(syst)}) \times 10^{-8}$, similar to that of the E391a result. 
After selecting events with two isolated clusters in the CsI fiducial region in time of each other, and in anti-coincidence with signals in the veto detectors,
the major remaining background comes from events where a halo neutron generates two hadronic clusters in the CsI. This background is modeled using special calibration runs with an Aluminum plate inserted in the $\kl$ beam path. After training two neural networks (NN) to select between photon and hadronic clusters on the basis of multiple cluster kinematic and shape variables, KOTO is able to reduce the neutron contamination to 0.18 $\pm$ 0.15 events, for a total background prediction inside the signal box of 0.36 $\pm$ 0.16 events.

\begin{figure}[htb]
\centering
\includegraphics[height=3.5in]{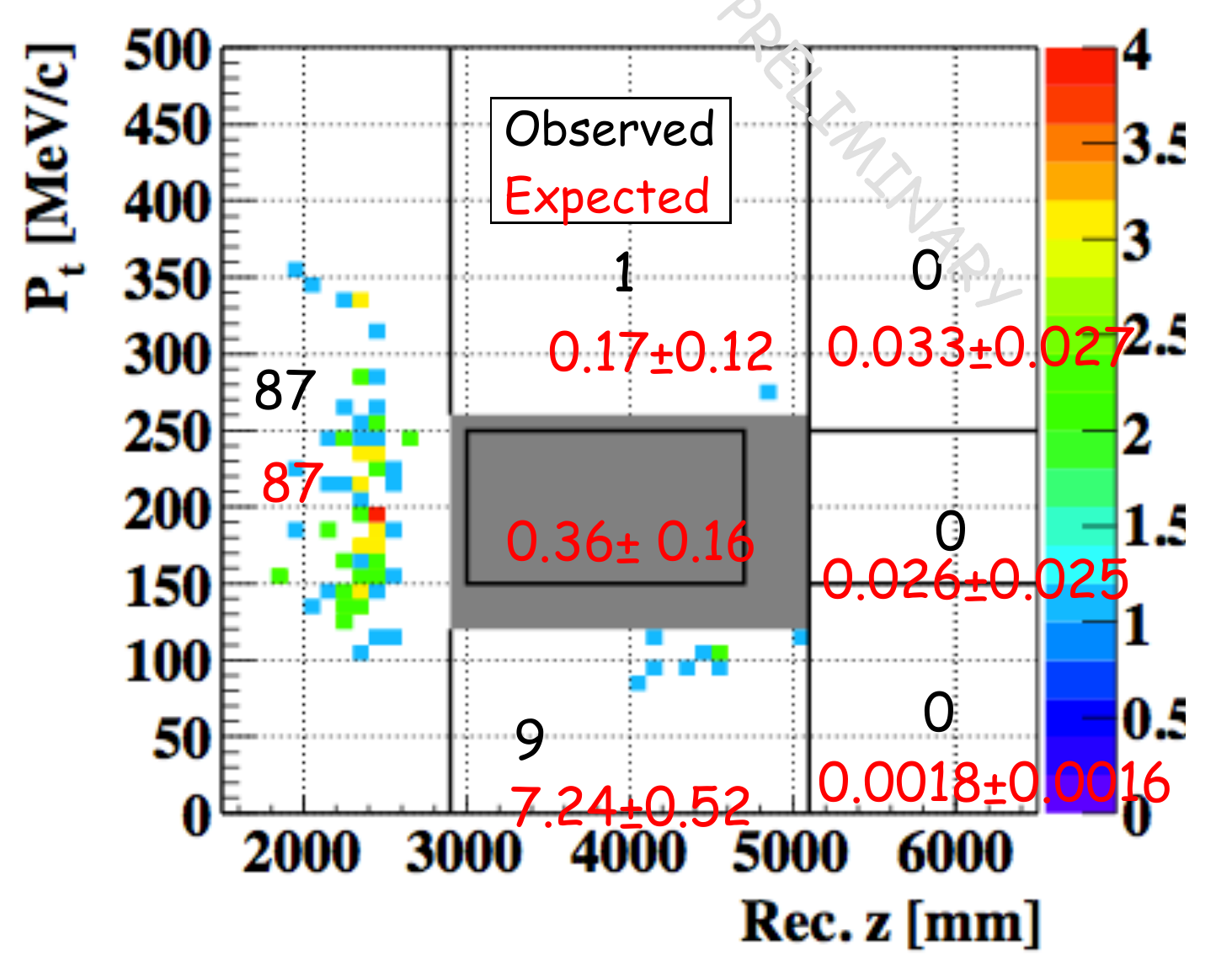}
\caption{Neutral pion P$_t$ vs reconstructed vertex for KOTO candidate events with a single fully reconstructed  $\pi^0$ passing all the cluster kinematic and shape variable NN selection. The rectangle inside the greyed area is the $\klpinn$ signal box. Good agreement is observed between data (black or top of two number) and MC expectations (red or bottom of two numbers) in all of the side-band regions.}
\label{fig:signal_box}
\end{figure}

KOTO pursued a blind analysis. Figure~\ref{fig:signal_box} shows the distribution of single $\pi^0$ candidate events in the (Z$_{vtx}$, P$_t$) plane: the signal box is defined as the region where
the $\pi^0$ vertex is reconstructed well inside the detector fiducial volume 
(3000~mm $<$ Z$_{vtx}  <$ 4800~mm) and with high transverse momentum (150~MeV/c$^2 < $ p$_T <$250  MeV/c$^2$).
Events in the low reconstructed vertex regions are due to halo neutron interactions with the upstream detector material generating a single $\pi^0$.
Events in the low P$_t$ region are due to $\rm{K}_L \rightarrow \pi^+ \pi^- \pi^0$ decays where the two charged pions escape down the beam pipe and interact with the beam pipe material in a region without any veto instrumentation.
Finally events in the high P$_t$ and high reconstructed vertex region are due to single halo neutrons generating two hadronic clusters in the CsI calorimeter.  After finding good agreement between data and Monte Carlo (MC) predictions in the side-bands of the signal region,  KOTO unblinds the data and finds 1 event in the signal box, for a preliminary limit B($\klpinn) < 5.0 \times 10^{-8}$ using Poission statistics.  Without the NN cuts, two events are found inside the signal box, consistent with the background prediction of 2.11 $\pm 1.06$ events. 

KOTO will restart taking data in Jan.~2015 with the goal of reaching the Grossman-Nir sensitivity, that is the indirect limit on B($\klpinn$) derived from B($\kcharged$) using the Grossman-Nir 
relation~\cite{GN_relation}: B($\klpinn$) $< 4.4$  B($\kcharged$). This relation is derived from isospin symmetry arguments and holds valid in any NP scenario. To suppress the background, the KOTO collaboration is actively looking into improving the veto coverage as well as honing the analysis tools needed to separate photon from hadronic clusters in the CsI calorimeter.

\subsection{The NA62 Experiment}

The NA62 experiment at the SPS 400 GeV/c proton beam at CERN has the goal of observing 100 $\kcharged$ events over 2 years. Unlike E787/E949, is does not rely on the detection of stopped kaons but rather on the detection of K$^+$ decays in flight using a 75 GeV/c unseparated hadron beam with 6\% kaon content. This will provide 4.8$\times 10^{-12}$ K$^+$ per year and allow NA62 to reach a S.E.S. $\approx 10^{-12}$. This very challenging experiment relies on separating the signal from the background by cutting on the missing mass of the neutrinos, 
$\rm{M}^2_{miss} = (\rm{P}_K - \rm{P}_{\pi})^2$ where the P$_K$ and P$_{\pi}$ are the 4-momenta of the charged kaon and pion, respectively. 

\begin{figure}[htb]
\centering
\includegraphics[height=2.5in]{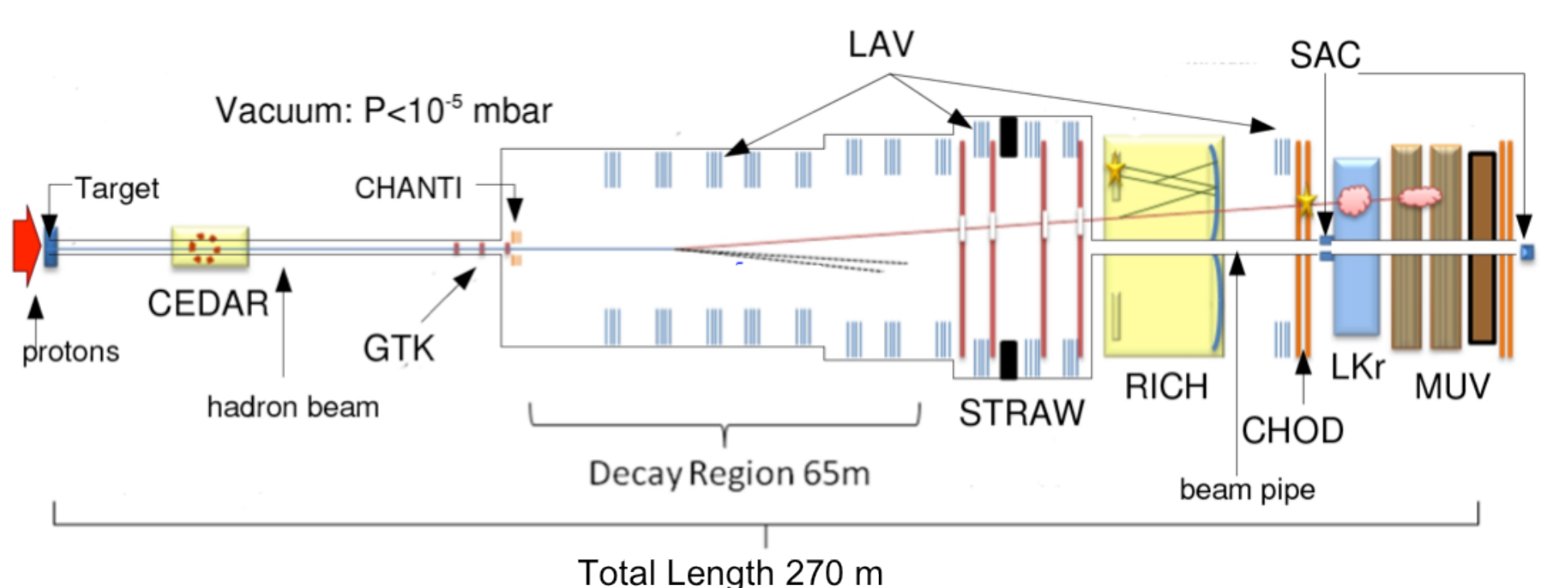}
\caption{Schematic view of the NA62 detector: the detectors are not drawn to scale.}
\label{fig:NA62}
\end{figure}

The detector, shown in Figure~\ref{fig:NA62}, measures the K and pion momentum with high resolution in low material trackers (GTK and STRAW) placed in vacuum at the entrance and exit of the 65m long decay region. Background rejection is achieved by a combination of photon veto detectors (LAV and SAC) and particle identification detectors with 10$^{-3}$ $\pi$/$\mu$ separation (RICH, LKr and MUV). A differential Cherenkov detector (CEDAR) upstream of the decay region provides kaon identification in the unseparated beam.

NA62 will have its first commissioning run from October to December 2014 with low beam intensity, during which it aspires to reach SM sensitivity.  Nominal intensity runs will follow in 2015, 2016 and 2017, before the LHC shutdown for the high luminosity upgrade. 

\section{$\kpimumu$ from NA48/2}\label{sec:kpimumu}

The $\kpimumu$ decay is a LNV decay mediated by a Maiorana neutrino. In some NP models, the rate is close to the experimental limit of 
B($\kpimumu) < 3\times 10^{-9}\,@\,90\%$ CL, obtained by the E865 experiment at the AGS in Brookhaven~\cite{E865_2000}.  The NA48/2 experiment has searched for this decay with data collected between 2003 and 2004 with a K$^+ + \rm{K}^-$ beam produced by 400 GeV/c primary SPS protons hitting a beryllium target .
The $\kpimumu$ rate is measured relative to the $\rm{K}^{\pm} \rightarrow \pi^{\pm}\pi^+\pi^-$ normalization sample. Given the $\pi$ and $\mu$  closeness in mass, the two samples are selected using the same trigger and similar identification cuts, thus allowing for cancellation of many systematic effects.

\begin{figure}[htb]
\centering
\includegraphics[height=2.5in]{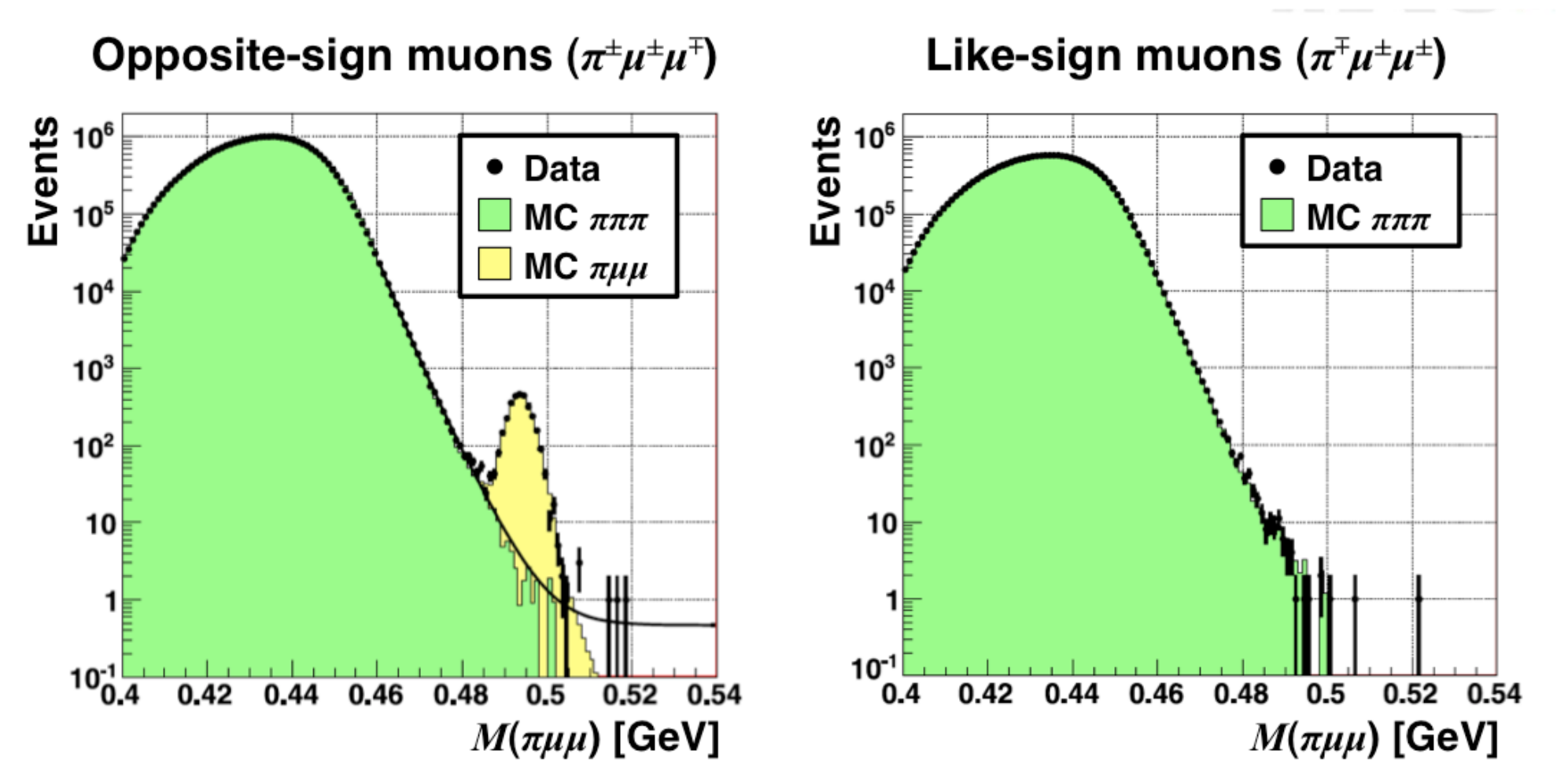}
\caption{Invariant mass in the $\pi\mu\mu$ hypothesis for NA48/2 candidate events with three charged tracks, one of which is identified as a pion: the left plot is for events with opposite-sign muons; the right plot is for events with like-sign muons. The green darker area is for $\rm{K}^{\pm} \rightarrow \pi^{\pm}\pi^+\pi^-$ MC prediction. The yellow lighter area is for $\rm{K}^{\pm} \rightarrow \pi^{\pm}\mu^+\mu^-$ MC prediction.}
\label{fig:kpimumu}
\end{figure}

Events with three well identified tracks coming from a common vertex are selected. For the  
$\pi\mu\mu$ sample, particle identification cuts are applied as to select events with one pion track and two muon tracks.  For the $3\pi$ sample, only the presence of one pion track is required.
Figure~\ref{fig:kpimumu} shows the invariant mass of the 3-track events in the $\pi\mu\mu$ 
hypothesis. While there is a clear peak around the K$^+$ mass for events with opposite sign muons,
the number of events with like-sign muons (52) is consistent with the background expectation 
(52.6$\pm$19.8) in the region 
$| \rm{M}_{\pi\mu\mu} -\rm{ M}_K | < 8$ GeV. Based on this number of candidate and background events, a limit of B($\kpimumu) < 1.1 \times 10^{-9}\,@\,90\%$ CL was obtained~\cite{NA48_2011}.  

\section{Rare Muon Decays}\label{sec:raremuon}

While LFV is well established for neutrinos, no such evidence exist for the charged leptons.
For CLFV to exist, neutrino must be Maiorana and not Dirac particles. But the presence of neutrino oscillations allow for $\mue$ conversion decays via higher 
order dipole penguin diagrams.  The SM prediction for this decay is beyond any measurable level because it is proportional to (m$_\nu$/m$_W$)$^4$ or $\approx 10^{-55}$.
This is a typical example of  a search for a highly improbable final state where any detection will provide an unambiguous sign of NP. There are many NP models predicting measurable rates: 
see~\cite{mue_review} and references therein for a recent review. 

\begin{figure}[htb]
\centering
\includegraphics[height=4.0in]{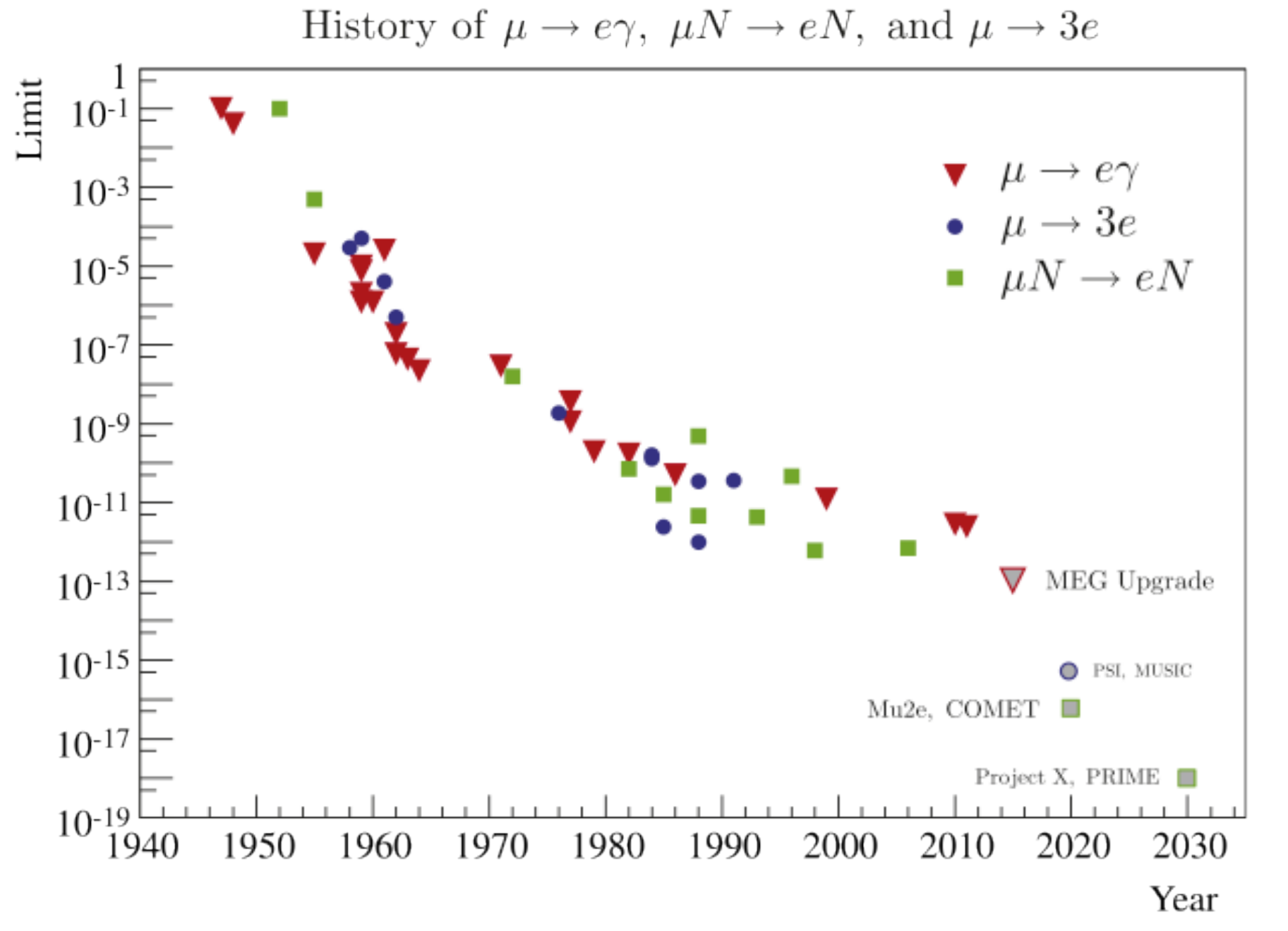}
\caption{Limit on $\mu \rightarrow e$ search through the years: the last points are the design limits for future proposed experiments. Courtesy of R.H. Bernstein, P.S. Cooper~\cite{mue_review}.}
\label{fig:mu_searches}
\end{figure}

Rare muon decay searches have a long history in three main channels, as shown in Figure~\ref{fig:mu_searches}:  $\meg$ searches; muon conversion in presence of a nucleon $\muN$; and $\mu \rightarrow 3\gamma$ searches. In the following, the present experimental situation for the first two is reported.

\subsection{The MEG Experiment}

The MEG experiment at PSI 1.3MW Proton Cyclotron searches for the $\meg$ decay. The signal has a clear two-body signal topology, with the production of simultaneous back-to-back positron-photon pairs.  The two main background contributions are from
inner brehmsstralung radiative muon decays, where the muon decays in
 $\meg\nu_e \bar{\nu_\mu}$ with low momentum neutrinos; and the much more dangerous (90\% of total contamination) accidental overlap of a photon from either 
 $e^+ e^- \rightarrow \gamma\gamma$ annihilation or from scattering of electrons off a nucleus, 
 $eN \rightarrow eN \gamma$, with the copious Michel decay 
 $\mu^+ \rightarrow e^+ \nu_e \bar{\nu_\mu}$.

Let's note how positive muons are preferred over negative muon partly because they are more easy to generate in high intensity beams, due to the larger rate of $\pi^+$ over  $\pi^-$ production in proton collisions and their consequent decay in flight. But mostly because $\mu^+$'s cannot be capture by nuclei, with subsequent release of neutrinos and photons which become source of accidentals.

The MEG detector envelops a thin target where muons from a high intensity DC muon beam are stopped. The decay products are subsequently detected using a high rate e$^+$ spectrometer consisting of a drift chamber and tracking counters. These trackers are immersed in a gradient magnetic field tuned to sweep out the Michel positrons while bending the monochromatic positrons from muon conversion with a projected constant radius, independently of the emission angle. 
A high resolution liquid xenon scintillation detector is optimized for the observation of the
back-to-back low energy $\gamma$ with  good efficiency, timing and spatial resolution in a relatively compact space.  

Using data up to 2011, MEG has published a limit B$(\meg) < 5.7\times10^{-13}\,@\,90\%$ CL~
\cite{meg_2011}. MEG has already collected more than twice the statistics used for this result and expects to reach a final sensitivity of 5$\times 10^{-13}$. An upgrade to the present detector has been proposed~\cite{MEGII}, with larger acceptance and better resolution, for a ultimate goal  
of 6$\times 10^{-14}$ sensitivity.

\subsection{The Mu2e Experiment}

The Mu2e experiment~\cite{Mu2e} will search for muon conversion in presence of an Al nucleus 
using a 8 GeV proton beam of $\approx$23 kW at FNAL. The experimental technique is to measure the number of $\meg$ conversions relative to muon captures, since details of the nuclear wave function cancel in the ratio.  Muons are "stopped" in the 1s state of the Al target nucleus and emit X-rays with a characteristic spectrum. These muonic atoms $\mu$'s can subsequently undergo three different decays: nuclear capture 
$\mu^- +$ N(A,Z) $\rightarrow \nu_\mu + $N(A+1,Z-1) by a nucleus N of atomic mass A and atomic number Z, which happens in 61\% of the cases for Al nuclei; decay in orbit (DIO) $\mu^- \rightarrow e^- \bar{\nu_e} \nu_\mu$, which happens in the remaining 39\% of the cases; and conversion to an electron,  $\mu^- + N(A,Z) \rightarrow e^- + N(A,Z)$, whose rate Mu2e aims to measure. 
 
 The experimental signature is a monochromatic electron of energy 
 $E_{\mu e}  = m_\mu - E_b - E_\mu^2/2m_N$, where $E_b$ is the muonic binding energy,  proportional to $Z^2$ and well known. For Al (Z=13), $E_{\mu e} = 104.973$ MeV,
completely separate from the 52.8 MeV endpoint of the Michel spectrum of electrons from free muons. But the Michel spectrum of electrons from DIOs is modified by the presence of the atomic nucleus momentum. In  particular the sharp energy endpoint is stretched to energies up to $E_{\mu e}$, pushing Michel electrons into the signal window. This happens with a rate of $10^{-17}$ within 1 MeV of $E_{\mu e}$, which limits the maximum sensitivity of the experiment~\footnote{An experiment with similar design and sensitivity, COMET, has also been proposed at a 8~GeV proton beam of intensity up to 56 kW in J-PARC~\cite{COMET}.}. This will still eclipse the present limit of $B(\meg) < 5.7 \times 10^{-13}\,@\,90\%$ CL.  set by the SINDRUM-II experiment at PSI~\cite{SINDRUM}. 

To reach its design sensitivity, Mu2e must ensure an energy resolution below 1 MeV and minimal energy losses for e$^-$.  The detector is places at the end of an S-shaped solenoid which selects negative pions produced by protons hitting a tungsten target. The decay in flight of $\pi^-$'s generate the slow muons to be stopped in the Al target. The S-shape eliminates line-of-sight photons and other neutral able to fake the electron signal. The detector itself consists of a tracker plus calorimeter in a solenoidal field at a radius large enough to see only particles with large transverse momentum. Main backgrounds are from cosmic muons producing an electron inside the signal energy window in the stopping target; and converting photons from radiative pion captures. The first is controlled by using a highly efficient cosmic veto; the second by using a pulsed beam which allows to separate in time the decay products of the faster pion capture vs the decay products of the muon capture. The first physic run for the Mu2e experiment is planned for 2019.

\section{Conclusions}

This paper reports on the status of selected rare decays of kaons and muons.
Recents results on B($\klpinn$) from KOTO, B($\kpimumu$) from NA48/2  and B($\meg$) from MEG are presented. The design sensitivity for B($\kcharged$) from NA62 and B($\muN$) from Mu2e are also discussed.



\Acknowledgements
I want to thank the organizers of PIC2014, especially Sabine Lammers, for the invitation to a very well organized and stimulating conference.


\begin{thebibliography}{99}


\bibitem{Brod_2011}
J.~Brod, M.~Gorbhan, E.~Stamou, PRD 83, 0340030 (2011).

\bibitem{E787_2009}
A.V. Artomonov et al., Phys. Rev. D 79, 092004 (2009).

\bibitem{E391a_2010}
J.K. Ahn et al., Phys. Rev. D 81, 072004 (2010).

 \bibitem{Straub_2011}
 D.M.Straub, arXiv:1012.3893v2 [hep-ph].

\bibitem{koto}
J.Comfort et al., {\it Proposal for $\klpinn$ Experiment at J-Parc"} (2006). Retrieved from {\tt http://koto.kek.jp/}

\bibitem{GN_relation}
Y. Grossman and Y. Nir, Phys. Lett. B 398, 163 (1997) [hep-ph/9701313].

\bibitem{E865_2000}
R.Appel et al, PRL 85, 2877 (2000).	

\bibitem{NA48_2011}
J.R. Batley et al., NA48/2 collaboration, Phys Lett B 697, 107 (2011).

\bibitem{mue_review}
R.H. Bernstein, P.S. Cooper, Physics Reports 532, 27Ð64 (2013).

\bibitem{meg_2011}
J. Adam et al., Phys. Rev. Lett 110, 201801 (2013)

\bibitem{MEGII}
A.M.Baldini et al., arXiv:1301.7225 [physics.ins-det]

\bibitem{Mu2e}
R. Abrams et al., arXiv:1211.7019 [physics.ins-det].

\bibitem{SINDRUM}
W.H. Bertl et al.,  Eur.Phys.J. C47 337-346 (2006).

\bibitem{COMET}
D.Bryman et al., {\it An Experimental Search for Lepton Flavor Violating $\mu^-$ - e$^-$ Conversion
at Sensitivity of 10$^{?16}$ with a Slow-Extracted Bunched Proton Beam"} (2006). Retrieved from
{\tt j-parc.jp/researcher/Hadron/en/pac\_1207/pdf/E21\_2012-10.pdf}.

\end{thebibliography}
\end{document}